\documentclass[prl,aps,superscriptaddress,floatfix,nofootinbib,twocolumn]{revtex4}
\usepackage{exscale}                  % correct scaling of math-symbols
\usepackage[intlimits]{amsmath}       % improved mathematical typesetting
\usepackage{amsfonts}
\usepackage{amssymb,amscd}
\usepackage[dvips]{epsfig}                   
\usepackage{array}
\usepackage{wrapfig}
\usepackage{color}
\newcommand{\beqa}{\begin{eqnarray}}
\newcommand{\eeqa}{\end{eqnarray}}
\newcommand{\beq}{\begin{equation}}
\newcommand{\eeq}{\end{equation}}

\definecolor{myred}{rgb}{1,0.8,0.8}

\begin{document}
\title{Analytic structure of the Landau gauge gluon propagator}
\author{Stefan Strauss}
\affiliation{Institut f\"ur Theoretische Physik,
  Justus-Liebig-Universit\"at Gie\ss{}en,
  Heinrich-Buff-Ring 16,
  D-35392 Gie\ss{}en, Germany}
\author{Christian~S.~Fischer}
\affiliation{Institut f\"ur Theoretische Physik,
  Justus-Liebig-Universit\"at Gie\ss{}en,
  Heinrich-Buff-Ring 16,
  D-35392 Gie\ss{}en, Germany}
\author{Christian Kellermann}
\affiliation{Institut f\"ur Kernphysik, 
 Technische Universit\"at Darmstadt,
 Schlossgartenstra{\ss}e 9, 64289 Darmstadt}
\date{\today}
\begin{abstract}
The analytic structure of the non-perturbative gluon propagator contains 
information on the absence of gluons from the physical spectrum of the 
theory. We study this structure from numerical solutions in the complex 
momentum plane of the gluon and ghost Dyson-Schwinger equations in Landau 
gauge Yang-Mills theory. The resulting ghost and gluon propagators are
analytic apart from a distinct cut structure on the real, timelike
momentum axis. The propagator violates the Osterwalder-Schrader positivity 
condition, confirming the absence of gluons from the asymptotic 
spectrum of the theory.
\end{abstract} 

\maketitle

{\bf Introduction}\\
One of the fundamental properties of QCD is the absence of its elementary
degrees of freedom, the quarks and gluons, from the physical spectrum of 
the theory. The associated problem of quark confinement is a much debated
issue \cite{Greensite:2011zz}. In this discussion 
it is useful to distinguish between two notions of confinement. 
One is in terms of color confinement, i.e. the absence of colored states from the 
asymptotic, physical state space of the theory. The other is strictly related 
to the center symmetry of Yang-Mills theory. Both notions are not 
equivalent. If center symmetry is unbroken, there exists a linear rising 
potential between static color charges in the fundamental representation of 
the gauge group, stretching out to arbitrary distances $r$ of the charges. 
In QCD, the mere presence of fundamental dynamical charges breaks this symmetry. 
Consequently, string breaking sets in at a sufficiently large 
separation $R$ of the fundamental test charges and the potential becomes 
flat for $r>R$. Thus, if confinement is defined in terms of unbroken center 
symmetry, QCD is not a confining theory \cite{Greensite:2011zz}. 
Gluons are also not confined in this sense, since they life in the adjoint
representation of the gauge group. When two gluons are separated far enough
from each other they pull two additional gluons out of the vacuum and dress up 
to form colorless bound states. This brings us back to the other notion of 
confinement: the absence of colored asymptotic states. A possible explanation 
of this absence are positivity violations. By definition, the asymptotic Hilbert 
space ${\cal H}_{phys}$ of colorless physical particles has to be positive 
(semi-)
definite, otherwise a probabilistic interpretation of its S-matrix 
elements would not be possible. Thus, positivity violations in the gluon 
propagator constitute a sufficient signal for the absence of gluons from 
the asymptotic spectrum of the theory. 

Another source of interest in the analytic structure of the gluon 
propagator comes from Heavy-Ion collisions. Currently, there is great 
activity both from theory and experiment at RHIC and ALICE/LHC to 
shed light on the properties of the quark-gluon plasma (QGP), i.e. strongly
interacting matter at large temperatures and/or density. Transport models 
like the Parton-Hadron-String Dynamics approach (PHSD) \cite{Cassing:2009vt} 
analyze the dynamics of quarks and gluons in the QGP. An important input 
into these calculations is the temperature dependent spectral function of 
the gluon, a quantity directly related to its analytical structure.
Whether and how this structure changes below and above the deconfinement
transition is currently an open question \cite{Maas:2011se}. 

How can the zero temperature analytic structure of the gluon propagator look 
like? Based on studies of the gauge fixing problem, Gribov \cite{Gribov:1977wm} 
and later on Zwanziger \cite{Zwanziger:1989mf} suggested a form for the gluon 
propagator with complex conjugate poles at purely imaginary squared Euclidean 
momenta. A generalization with complex conjugate poles in the negative half-plane of 
squared complex momenta has been suggested by Stingl in Ref.~\cite{Stingl:1985hx} 
and has been recently explored in detail in the Refined Gribov-Zwanziger
framework Ref.~\cite{Cucchieri:2011ig}. These forms have in common that they may 
pose a problem for the analytic continuation of the theory from Minkowski space to 
the Euclidean formulation used by lattice gauge theory and functional methods. 
Furthermore, in their pure form they do not account for the perturbative,
logarithmic running of the propagator in the large momentum region. An alternative 
form with a branch cut structure for real and time-like squared momenta has 
been proposed in Ref.~\cite{Alkofer:2003jj} and found to compare well with 
numerical results for the propagator and its Schwinger functions in the 
Dyson-Schwinger (DSE) approach. 

All explicit calculations of the gluon propagator so far have been restricted to the 
real and spacelike Euclidean momentum domain. Clearly, in order to pin down the analytic 
structure of the gluon propagator, an extension of these calculations into the complex 
momentum domain is highly desirable. In this letter we report the first results of 
such a calculation. Within the continuum formulation of Landau gauge Yang-Mills theory 
we solved the coupled system of Dyson-Schwinger equations for the non-perturbative gluon 
and ghost propagators in the complex momentum plane and extract the analytic structure 
at time-like momenta. As a main result we find analytic propagators everywhere, apart 
from cuts on the real, timelike momentum axis. 

{\bf The framework}\\
The Dyson-Schwinger equations (DSEs) for the ghost and gluon propagators are 
%%%%%%%%%%%%%%%%%%%%%%%%%%%%%%%%%%%%%%%%%%%%%%%%%%%%%%%%%%%%%%%%%%%%%%%%%%%%
\begin{figure}[t]
\centerline{\includegraphics[width=0.4\textwidth]{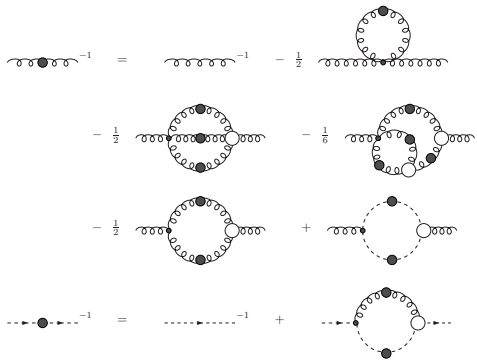}}
\caption{Dyson-Schwinger equations for the gluon and
  ghost propagator. Filled circles denote dressed propagators and
  empty circles denote dressed vertex functions.}
\label{fig:DSE}
\end{figure}
%%%%%%%%%%%%%%%%%%%%%%%%%%%%%%%%%%%%%%%%%%%%%%%%%%%%%%%%%%%%%%%%%%%%%%%%%%%%
shown in Fig.~\ref{fig:DSE}. They form a coupled set of integral equations
with renormalized, bare and dressed propagators and vertices.  
In Landau gauge, the explicit form of the propagators is given by 
\begin{eqnarray}\nonumber 
 D_{\mu \nu}(p) &=& \left(\delta_{\mu \nu} -
    \frac{p_\mu p_\nu}{p^2}\right) \frac{Z(p^2)}{p^2}\\ 
    D_G(p) &=& -\frac{G(p^2)}{p^2}
  \label{props} 
\end{eqnarray}
with the gluon dressing function $Z(p^2)$ and the ghost dressing function
$G(p^2)$. These functions can be numerically determined from their DSEs 
provided explicit expressions for the dressed ghost-gluon, three-gluon and 
four-gluon vertices are known. Since these satisfy their own DSEs containing
unknown higher n-point functions, in practice one needs to truncate this tower
to generate a closed and solvable system of equations. Certainly, meaningful
results can only be achieved by careful control of the quality of such a 
truncation. 
%%%%%%%%%%%%%%%%%%%%%%%%%%%%%%%%%%%%%%%%%%%%%%%%%%%%%%%%%%%%%%%%%%%%%%%%%%%
\begin{figure}[t]
\centerline{\includegraphics[width=0.4\textwidth]{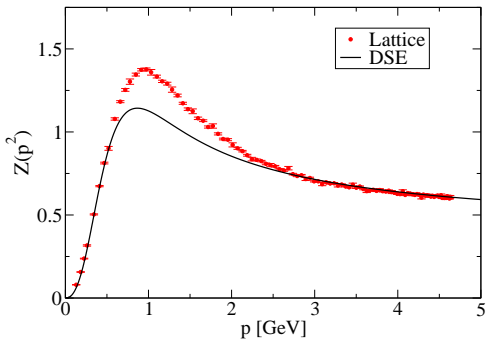}}
\caption{Results for the gluon dressing function $Z(p^2)$
  from lattice calculations \cite{sternbeck06} compared to the result 
  from DSEs \cite{Fischer:2008uz}.}
\label{fig:real}
\end{figure}
%%%%%%%%%%%%%%%%%%%%%%%%%%%%%%%%%%%%%%%%%%%%%%%%%%%%%%%%%%%%%%%%%%%%%%%%%%%%

A scheme which maintains multiplicative renormalizability and 
transversality has been devised in Ref.~\cite{Fischer:2008uz}; for transverse 
projection of the gluon DSE this scheme is equivalent to the one used previously 
in Ref.~\cite{Fischer:2002hna}. It uses a bare ghost-gluon vertex, a choice which 
is close to the results of corresponding lattice calculations \cite{Cucchieri:2008qm},
an ansatz for the dressed three-gluon vertex in terms of the propagator functions,
see \cite{Fischer:2002hna,Fischer:2008uz} for details, and a vanishing four-gluon
interaction. Given this choice, the coupled system of DSEs can be solved numerically.
The resulting solution for the gluon dressing function $Z(p^2)$ has been discussed
in Refs.~\cite{Fischer:2002hna,Fischer:2008uz} and, for the convenience of the 
readers, is shown again in Fig.~\ref{fig:real} together with corresponding lattice 
results \cite{sternbeck06}. In the large momentum region, where the perturbative
behavior sets in, both approaches agree very well. This is also true in the low
momentum region. In the deep infrared, the gluon dressing function displays the 'massive' behavior
$Z(p^2) \sim p^2$. Such 'decoupling' type of solutions, as opposed to 
'scaling' \cite{Watson:2001yv}, have been suggested long ago
\cite{Cornwall:1981zr} and have been revived in Refs.~\cite{Aguilar:2008xm,Boucaud:2008ji,Dudal:2008sp}.
Large volume lattice results agree with this type of solutions \cite{Cucchieri:2008fc},
although it remains a matter of current debate whether problems with gauge
fixing in the context of Gribov copies are already well under control 
\cite{vonSmekal:2008ws,Sternbeck:2008mv,Cucchieri:2009zt,Maas:2009se}. 

In the 
mid-momentum region around one GeV there are differences between the DSE and the 
lattice result on the twenty percent level which have to be attributed to the 
above discussed vertex truncations for the ghost-gluon, three-gluon and four-gluon 
vertex. Improvements for the dressed ghost-gluon vertex
have been discussed in Refs.~\cite{Boucaud:2011eh,Pennington:2011xs}. Furthermore, 
first studies of other types of ansaetze for the dressed three-gluon vertex are 
available \cite{Pennington:2011xs} and studies in the background gauge Pinch-technique 
scheme emphasize the importance of poles in the longitudinal parts of the three-gluon 
vertex, see \cite{Binosi:2009qm} for a review. While all these studies are interesting 
on systematic grounds, the resulting solutions for the gluon dressing function are
not closer to the lattice result than the one shown in Fig.~\ref{fig:real}. The remaining
difference may therefore very well be attributed to the missing two-loop diagrams
involving the four-gluon interaction. Indeed, pointwise agreement with the lattice data 
has been achieved within the framework of functional renormalisation group 
equations \cite{Fischer:2008uz}, where such contributions can be taken into account
due to the exact one-loop structure of the equations. Within the DSE framework the 
technical complications arising from the two-loop diagrams only allowed for 
phenomenological treatments of these contributions so far \cite{Bloch:2003yu}, and 
we therefore prefer to defer a study of the influence of these terms to future work.

The numerical techniques necessary to solve a coupled set of DSEs in the complex plane 
have been explored up to now only in the context of the fermion propagator, see \cite{Maris:1995ns} and 
the appendix of \cite{Alkofer:2003jj}. The basic idea is to shift the contour of 
radial integration in the loop integral into the complex plane such that singularities
in the angular integral are meliorated. For this work we adapted this method for the 
ghost and gluon system. Details will be given elsewhere. We have cross-checked our numerics 
also by employing a different method which solves the DSEs directly on a grid of complex 
momenta without any shifts in the integrals \cite{CK_diss}. The results of both 
methods agree well for a large range of complex momenta. However, close to the timelike 
momentum axis, our first method clearly delivered much more stable results and is therefore 
to be preferred.

\begin{figure}[t]
\centerline{\includegraphics[width=0.42\textwidth]{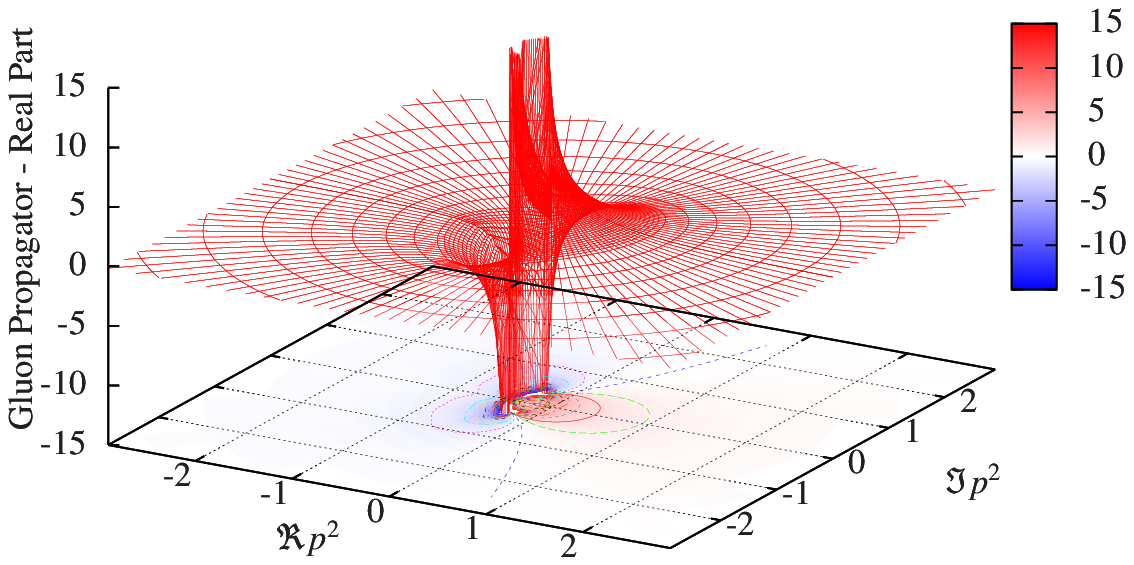}}
\centerline{\includegraphics[width=0.42\textwidth]{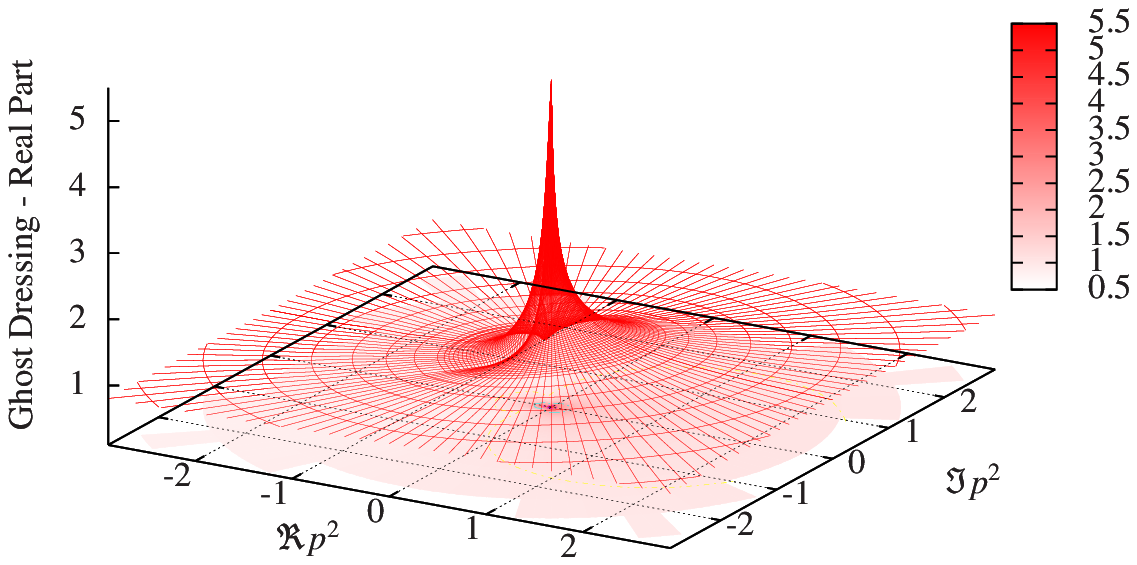}}
\caption{Results for the real part of the gluon propagator function $D(p^2)$
  and the ghost dressing function $G(p^2)$ in the complex momentum plane including 
colored contour maps and lines. The displayed range of the gluon propagator is restricted in order to 
resolve smaller structures. See text for the extrema of $\Re D(p^2)$.}
\label{fig:c_real}
\end{figure}

\begin{figure}[th]
\centerline{\includegraphics[width=0.42\textwidth]{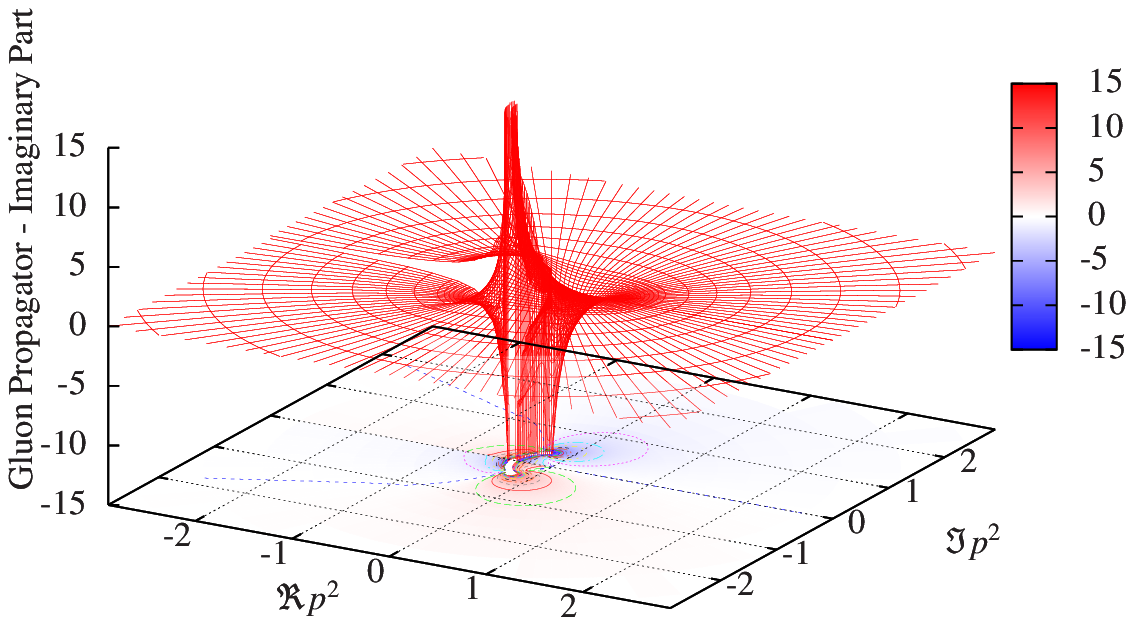}}
\centerline{\includegraphics[width=0.42\textwidth]{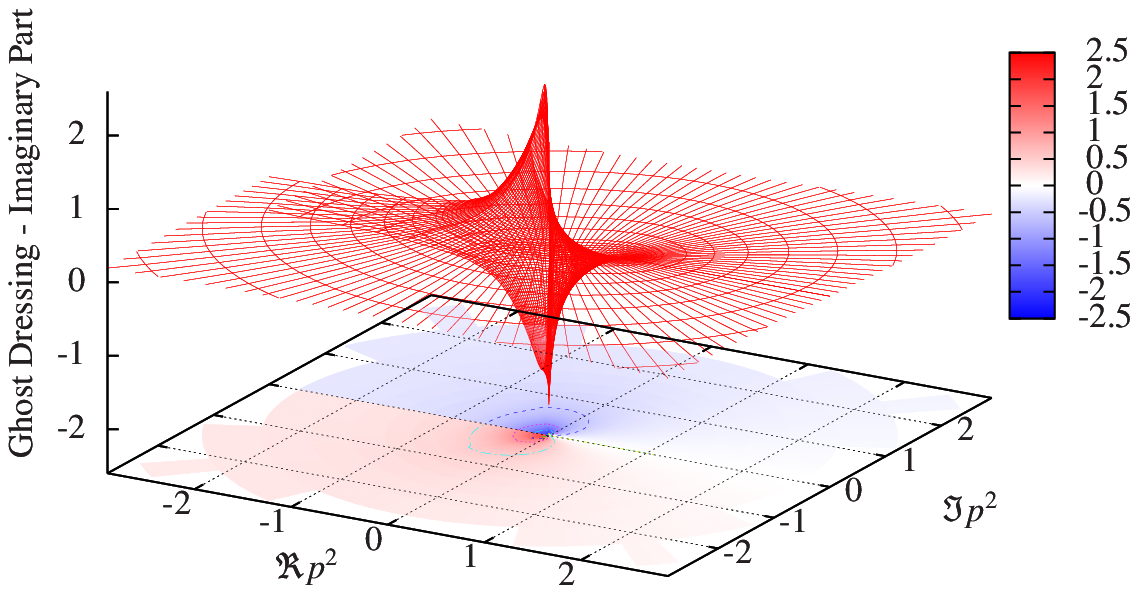}}
\caption{Results for the imaginary part of the gluon propagator function $D(p^2)$
  and the imaginary part of the ghost dressing function $G(p^2)$ in the complex momentum plane 
including colored contour maps and lines. The displayed range of the gluon propagator is restricted in order to 
resolve smaller structures. See also Fig.~\ref{fig:spectral} for the full scale.}
\label{fig:c_imag}
\end{figure}

{\bf Results and discussion}\\
Our results for the analytic structure of the gluon and ghost propagator in SU(N) Yang-Mills 
theory are shown in Figs.~\ref{fig:c_real} and \ref{fig:c_imag}.
%%%%%%%%%%%%%%%%%%%%%%%%%%%%%%%%%%%%%%%%%%%%%%%%%%%%%%%%%%%%%%%%%%%%%%%%%%%%
%%%%%%%%%%%%%%%%%%%%%%%%%%%%%%%%%%%%%%%%%%%%%%%%%%%%%%%%%%%%%%%%%%%%%%%%%%%%
%%%%%%%%%%%%%%%%%%%%%%%%%%%%%%%%%%%%%%%%%%%%%%%%%%%%%%%%%%%%%%%%%%%%%%%%%%%%
%%%%%%%%%%%%%%%%%%%%%%%%%%%%%%%%%%%%%%%%%%%%%%%%%%%%%%%%%%%%%%%%%%%%%%%%%%%%
Let us first discuss the real parts of the propagators. In the lower diagram of 
Fig.~\ref{fig:c_real}
we see a spike structure of the ghost dressing function, which is located
at the origin of the complex momentum plane. Since we have chosen a decoupling 
type of solution for the ghost-gluon system, the ghost dressing function is 
finite at $p^2=0$.
At our numerical infrared cutoff $|\epsilon^2|=10^{-5}$ the value $G(\epsilon^2)$ 
depends slightly on the direction from which zero is approached. In our calculation 
$G(\epsilon^2)=5$ when $p^2 \rightarrow 0^+$, but $G(-\epsilon^2)=5.005 \mp{\rm i}\, 0.004$ when 
$p^2 \rightarrow 0^-$ on the real axis. Thus, the real part of the ghost dressing function is 
almost symmetric around the origin of the complex momentum plane.
For the real part of the gluon propagator in the upper diagram of 
Fig.~\ref{fig:c_real} the situation is entirely different. Again,
the propagator is finite at $p^2=0$ but shows large positive values 
for complex momenta and negative structures close to the negative real 
momentum axis. The positive spikes extend up to $\Re D(p^2) = 40$ GeV$^2$ 
staying definitely finite. Closer to the negative real momentum axis the 
propagator becomes negative for $|p^2|$ larger than some finite value on the 
negative real momentum axis. The corresponding narrow dip is finite in depth 
and sizably extends $0.3$ GeV$^2$ out into the complex momentum plane. 
The minimum value of the dip is approximately $\Re D(p^2) = -105$ GeV$^2$. 

Now let us discuss the imaginary part in Fig.~\ref{fig:c_imag}. Here we 
clearly see a cut-structure
emerging for both, the gluon propagator and the ghost dressing function along 
the negative real momentum axis. No further structure is seen in the complex
momentum plane. We thus arrive at an important result of our study:
the ghost and gluon propagators have nontrivial analytic structure 
only on the timelike real momentum axis. This is in contrast to the expectations
from the studies of Gribov, Zwanziger and Stingl 
\cite{Gribov:1977wm,Zwanziger:1989mf,Stingl:1985hx,Cucchieri:2011ig}, which all 
assumed singularities away from the real momentum axis. We find no evidence
for these. Within the present numerical accuracy, the cuts are sharply peaked but 
finite. Whether an even more precise treatment leads a singularity in $\Im D(p^2)$, 
as assumed in the fits to the DSE results in Ref.~\cite{Alkofer:2003jj},
remains an open question.

%%%%%%%%%%%%%%%%%%%%%%%%%%%%%%%%%%%%%%%%%%%%%%%%%%%%%%%%%%%%%%%%%%%%%%%%%%%%
\begin{figure}[t]
\centerline{\includegraphics[width=0.5\textwidth]{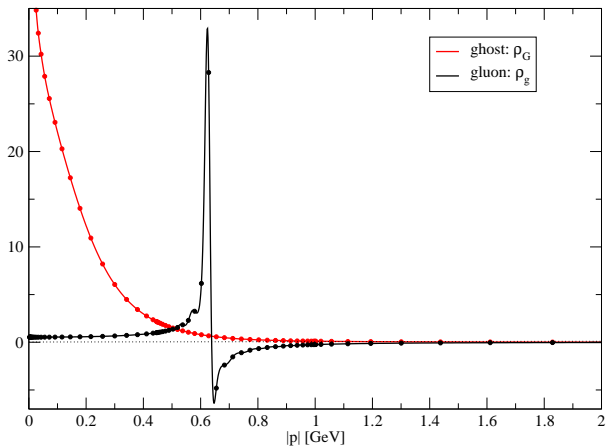}}
\caption{Results for the gluon spectral function and the ghost spectral 
function as a function of momentum.}
\label{fig:spectral}
\end{figure}
%%%%%%%%%%%%%%%%%%%%%%%%%%%%%%%%%%%%%%%%%%%%%%%%%%%%%%%%%%%%%%%%%%%%%%%%%%%%

The cuts in the imaginary part of the ghost and gluon propagators are directly 
related with the corresponding spectral functions,
\beq 
\rho_{G,g}(p^2) =  - \Im\{D_{G,g}(p^2)\}/\pi
\eeq
with $D_G(p^2)=-G(p^2)/p^2$, 
$D_g(p^2)=Z(p^2)/p^2$ and the momentum $p^2$ on the negative real momentum 
axis of the upper complex half plane. 
We therefore show them more closely
in Fig.~\ref{fig:spectral}. Our numerical results are obtained on a grid of 
momentum points which are displayed explicitly, whereas the interpolation is 
done via Chebychev polynomials and serves to guide the eye. For the ghost 
spectral function (and most of the gluon) this interpolation clearly works 
also on a quantitative level; however, the interpolation for the gluon in the region 
$0.5 < |p| < 0.7$ GeV can only be regarded as a qualitative one. 
Whereas the ghost spectral function is dominated by the massless $1/p^2$ pole 
in the propagator, the gluon is clearly different. Its imaginary part rises 
first, turns over and crosses through zero,
turns again and then approaches zero from below at large timelike momenta.
The exact locations and heights of the maxima in the positive and negative 
region cannot be determined precisely within the present numerical accuracy. 
Nevertheless, the qualitative behavior is
fixed from the explicitly calculated points shown in the plot. 
The gluon spectral function obeys the Oehme-Zimmermann normalization 
condition \cite{Oehme:1979ai}
\beq
Z_3^{-1}=\int\rho_g(s)\; {\rm d}s,
\eeq
where $Z_3$ denotes the gluon wave function renormalization constant, with 
a deviation of 10 percent. 
This provides a measure on the accuracy achieved in the present computation.
The negative contributions to the gluon spectral function indicate its 
absence from the asymptotic spectrum of the theory. In general, the cuts in 
the ghost and gluon propagators signal the radiation of unphysical 
particles (ghosts and gluons) from unphysical particles. Moreover, the gluon
is certainly not a massive particle in the usual sense. Nevertheless one may 
be tempted to define 
something like an "effective mass" $m_g$ for the gluon from the location of the 
positive peak in the spectral function. Within the present accuracy we find
\beq
600 \,\text{MeV}\, <\, m_{g}\, < 700 \,\text{MeV}\,.
\eeq
We stress again, however, that $m_{g}$ is not a measurable quantity;
strictly speaking, it is just the scale where positivity violations 
in the gluon set in.

In this work we presented the first non-perturbative solution of the gluon and 
ghost propagators in the complex momentum plane together with an extraction of
their respective spectral functions. We presented results for the decoupling case;
a comprehensive comparison with scaling will be given elsewhere.
Besides the considerable theoretical interest
in these functions they are also a necessary input into the calculations of glueball
masses within the framework of Bethe-Salpeter equations. Corresponding results will 
be detailed in a subsequent work. 

{\bf Acknowledgements}\\
We thank Reinhard Alkofer, Jan Pawlowski and Lorenz von Smekal for fruitful 
discussions. This work has been supported by the Helmholtz Young Investigator 
Grant VH-NG-332 and the Helmholtz International Center for FAIR within the 
LOEWE program of the State of Hesse.

\end{document}